# Polarization relaxation in thin-film relaxors compared to that in ferroelectrics

M. Tyunina, J. Levoska, and I. Jaakola

*Microelectronics and Materials Physics Laboratories, University of Oulu, PL 4500, FI-90014 Oulun Yliopisto, Finland*

Epitaxial thin films of relaxor $PbMg_{1/3}Nb_{2/3}O_3$ and $PbSc_{0.5}Nb_{0.5}O_3$, and ferroelectric $PbZr_{0.65}Ti_{0.35}O_3$, $Pb_{0.955}La_{0.045}Zr_{0.65}Ti_{0.35}O_3$, and $Ba_{0.4}Sr_{0.6}TiO_3$ were prepared, and their dielectric properties were studied in a broad range of the measurement conditions. In the ferroelectric state, the presence and the change of configuration of the domains determined both the dynamic dielectric nonlinearity and the polarization hysteresis. In thin-film relaxors, the orientation of the randomly interacting dipoles in a random field was responsible for the dynamic dielectric nonlinearity, while the observed hysteresis was suggested to arise due to connection between the applied field and the relaxation times of both the dipoles and the internal field. In thin-film $(Ba,Sr)TiO_3$, the high-temperature dielectric hysteresis was found to be relaxorlike.

## I. INTRODUCTION

In perovskite relaxor ferroelectrics (RFE), both the nature of glassy state and the mechanism of the relaxor to ferroelectric (FE) phase change are the subjects of intensive research.[1] Many of the features of RFE can be modeled as resulting from orientation of the randomly interacting dipoles embedded in a polarizable matrix, with a broad distribution of the local electric field, or random field.[2] The distribution of the potential barriers between different orientational states determines a broad distribution of relaxation times. However, both another type of nanodomain kinetics and coexistence of ferroelectric and glassy types of order have been experimentally detected.[3] Also a direct evidence of the existence and evolution of the nanodomains has been obtained.[4]

In thin-film RFE, the main features of RFE have been experimentally observed. Vogel-Fulcher relationship, deviation from the Curie-Weiss behavior, temperature evolution of the relaxation time spectra, temperature dependence of the glasslike order parameter, and empirical scaling of the dielectric peak have been found to be similar to those in bulk RFE.[5,6] However, also both the ac field dependent piezoelectric effect and the relatively broad polarization – electric field, or $P-E$, hysteresis loops have been observed.[7-9] A FE-like self-polarized state has been suggested to be responsible for that.[7] On the other side, both the onset of the polar state under an applied dc field[10] and the effects of applied ac field[6] have been demonstrated to result from the orientation of the randomly interacting dipoles in a random field. The presence of FE domains has not been detected.[11]

The purpose of this study was to clarify the possible contribution of FE-like domains to the polarization in thin-film RFE. The dynamic dielectric response determined by the FE domains[12,13] and that determined by the RFE dipoles[14,15] differ significantly, making it possible to distinguish the contributions. In thin-film heterostructures of various RFE and FE, the dielectric permittivity was measured as a function of frequency, temperature, amplitude of ac field, biasing dc field, and varying the field sweeping conditions. The presence of FE domains was demonstrated in thin films in FE state, while it was not detected in thin-film RFE. To explain the $P-E$ hysteresis and the dielectric hysteresis observed in thin-film RFE, the relaxation of both the dipoles and the internal field were considered. A good agreement between the suggested relaxation scenario and numerous experimental observations was found. Also the high-temperature dielectric hysteresis in thin-film $(Ba,Sr)TiO_3$ was qualitatively explained. The peculiarities of thin-film RFE compared to bulk were discussed.

## II. EXPERIMENT

### A. Thin-film heterostructures

Heterostructures of thin-film RFE or FE with a $La_{0.5}Sr_{0.5}CoO_3$ (LSCO) bottom electrode layer and Pt top electrodes were grown by *in situ* pulsed laser deposition on MgO (001) single crystal substrates.[9,16] The films of 300 nm thick $Pb_{0.955}La_{0.045}Zr_{0.65}Ti_{0.35}O_3$ (PLZT), 300 nm thick $PbZr_{0.65}Ti_{0.35}O_3$ (PZT), 250 nm thick $PbMg_{1/3}Nb_{2/3}O_3$ (PMN), 400 nm thick $PbSc_{0.5}Nb_{0.5}O_3$ (PSN), and 400 nm thick $Ba_{0.4}Sr_{0.6}TiO_3$ (BST) were prepared providing a set of different FE and RFE materials. At room temperature FE/RFE films were perovskite, pseudocubic, (001) oriented, with an in-plane epitaxial relationship FILM [100] || LSCO [100] || MgO [100]. More details on both the deposition procedure and the microstructure analysis can be found elsewhere.[9,16,17] The dielectric response of the heterostructures was measured along the [001] axis of the films, normal to the substrate surface. An HP 4284 *LCR* meter (frequency range of $10^2 - 10^6$ Hz) and an RT 6000 tester were used. The real part of the dielectric permittivity $\varepsilon'$ was determined as a function of frequency $f$ and amplitude $E_{AC}$ of the ac electric field in the temperature range of $T = 90 - 670$ K. Also both the magnitude $E_{DC}$ and the sweeping rate of the biasing dc electric field were varied.

### B. Dielectric nonlinearity

To distinguish the mechanisms of polarization relaxation – either due to presence of FE domains or due to orientation of the random dipoles – the dynamic dielectric nonlinearity was studied. In thin-film FE, the nonlinearity is known to originate from both reversible and irreversible processes, including domain wall motion and switching. The $P-E$ loops and a linear $\varepsilon'(E_{AC})$ dependence below the coercive field are the characteristics of FE state. In contrast in RFE, the behavior of $\varepsilon'(E_{AC})$ is more complex.[6] For RFE as an ensemble of the randomly interacting dipoles with a broad distribution of the local electric field, an analytical description of the dynamic properties is difficult, and it requires numerical simulations.[14,15] In particular, effects of an ac electric field have been simulated using the Monte-Carlo method.[15] A system of Ising-type dipoles, with randomly distributed interactions, and with thermally activated flipping of the local polarization, was considered:

$$H = -\sum_{i \neq j} \widetilde{J}_{ij}\sigma_i\sigma_j - E_{ext}\overline{\mu}\sum_i \frac{|\mu_i \cos\theta_i|}{\overline{\mu}}\sigma_i, \qquad (1)$$

where $\sigma_i, \sigma_j = \pm 1$ are dipole spins with $\sigma_i = 1$ ($\sigma_i = -1$) when the projection of the dipole moment $\vec{\mu}_i$ on the direction of the external field $\vec{E}_{ext}$ is positive (negative), $\overline{\mu}$ is the maximum magnitude of the dipole moments, and $\widetilde{J}_{ij}$ is the effective interaction energy between the nearest neighbor dipoles having a Gaussian distribution with width $\Delta J$. For the present study, the dynamic nonlinear dielectric permittivity $\Delta\varepsilon'$ was extracted from the results of the simulations.[15]

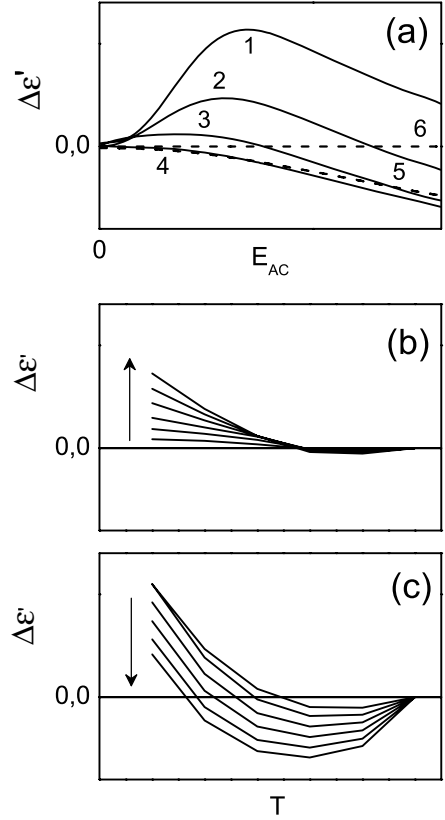

Fig. 1. The dynamic nonlinear dielectric permittivity $\Delta\varepsilon'$ extracted from the results of the simulations (a) as a function of amplitude $E_{AC}$ of ac electric field at different temperatures (the temperature corresponding to the curves 1 – 6 is increasing, $T_1 < T_2 < T_3 < T_4 < T_5 < T_6$); (b) as a function of temperature at smaller amplitudes $E_{AC}$ (arrow shows the direction of increase of $E_{AC}$); (c) as a function of temperature at large amplitudes $E_{AC}$ (arrow shows the direction of increase of $E_{AC}$).

The permittivity $\Delta\varepsilon'$ was determined as $\Delta\varepsilon' = \varepsilon'(E_{AC}) - \varepsilon'(E_{AC0})$, where the permittivity $\varepsilon'(E_{AC0})$ corresponded to the smallest amplitude $E_{AC}$ of the ac electric field. It was a non-monotonic function of the amplitude $E_{AC}$ [Fig. 1(a)]. The position of the maximum in $\Delta\varepsilon'(E_{AC})$ was moving to smaller fields with decreasing frequency. For smaller amplitudes $E_{AC}$, the permittivity $\Delta\varepsilon'$ increased on cooling [Fig. 1(b)], but for larger $E_{AC}$ it became a non-monotonic function of temperature [Fig. 1(c)]. The nonlinear $\Delta\varepsilon'$ was negative around the temperature of the dielectric maximum $T_m$ and it vanished at the Burns temperature, much higher than $T_m$. Such a specific evolution of $\Delta\varepsilon'$ has been recently detected in thin-film PMN.[6] Here the dynamic nonlinear $\Delta\varepsilon'$ was measured as a function of amplitude $E_{AC}$ at different temperatures and electric field $E_{DC}$, and compared with the results of the simulations.

## III. RESULTS

### A. Thin-film ferroelectrics

In both PZT and PLZT thin-film heterostructures, the real part of the dielectric permittivity $\varepsilon'$ measured on cooling at small ac electric field exhibited a broad frequency dependent peak [Fig. 2(a)]. The temperature $T_m$ of the dielectric maxima was frequency independent and rather close to that of ceramics.[9] At room temperature, the $P-E$ hysteresis loops were observed. This evidenced the occurrence of the para- to ferroelectric transition in the films. The frequency dispersion of $\varepsilon'$ was related to the film – electrode coupling.[18]

In FE state at room temperature, the permittivity $\varepsilon'$ was measured using sweeps of the biasing dc electric field superimposed with the small probing ac field [Figs. 2(b, c)]. It has been previously shown[12] that the integrated $\varepsilon'(E_{DC})$ curve could characterize a contribution to polarization from the reversible processes such as reversible displacements of the domain walls. Since each point of the $\varepsilon'(E_{DC})$ curve was characteristic for a certain domain configuration at a given field $E_{DC}$, also irreversible changes in the domain configuration and hence a normal ferroelectric hysteresis could be followed by the $\varepsilon'(E_{DC})$ curve. The observed dielectric hysteresis [Figs. 2(b, c)] was determined mainly by the switching of polarization. An asymmetry of the heterostructures and film – substrate mismatch[19] determined also the asymmetry of the $\varepsilon'(E_{DC})$ curves.

In FE films under subswitching conditions, the dependence of permittivity $\varepsilon'$ on the amplitude $E_{AC}$ has been found to be linear and hysteresis-free after the first sweep of $E_{AC}$.[13] Such a behavior of $\varepsilon'(E_{AC})$ has been interpreted in terms of the Rayleigh-type motion of the domain walls.[13] In the present study, the behavior of thin-film FE, both PZT and especially PLZT, had peculiarities compared to that observed in the relatively thick (> 1 μm) polycrystalline PZT films.[13] Similar to that in Ref. 13, the first sweep of ac amplitude was different from the consequent ones [Fig. 2(d)]. With increasing $E_{AC}$ a hysteresis-free increase of $\varepsilon'$ was detected for very small fields. In PLZT at $E_{AC} \geq 0.4$ MV/m, still below the coercive field, the $\varepsilon'(E_{AC})$ hysteresis loops [Fig. 2(e)] were found in the whole studied frequency range. In PZT, the hysteresis was less expressed but still noticeable at $E_{AC} \geq 1$ MV/m. This indicated a contribution from fast irreversible processes.

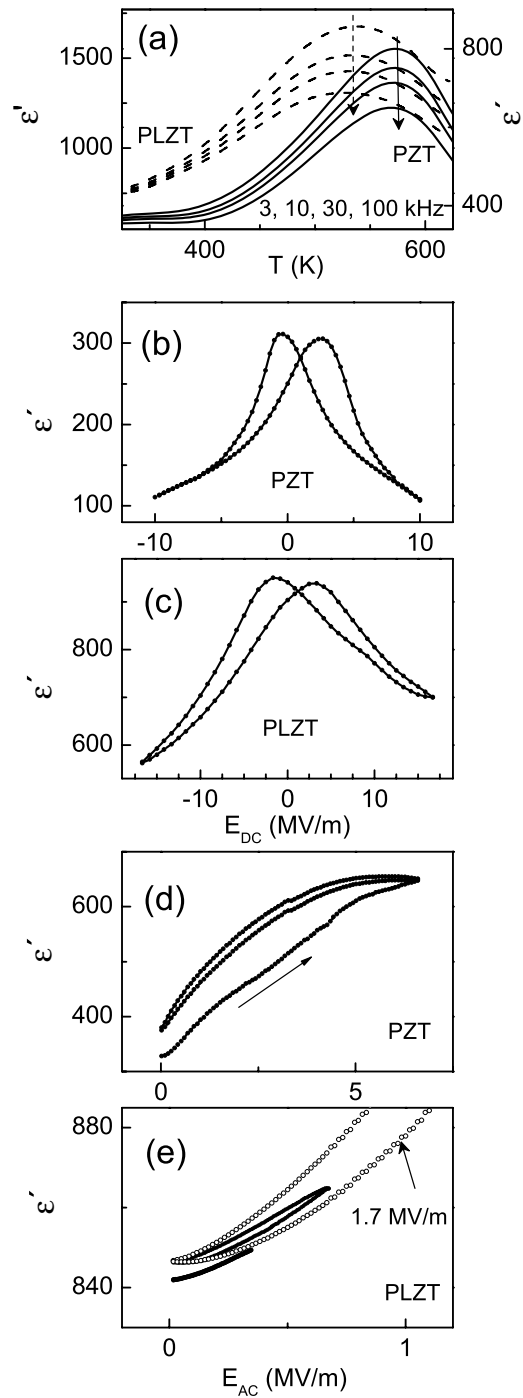

Fig. 2. The real part of the dielectric permittivity $\varepsilon'$ of thin-film FE (a) as a function of temperature $T$ measured on cooling in PZT and PLZT at different frequencies of small ac field (the arrows show the direction of increase of frequency); (b, c) as a function of applied dc electric field $E_{DC}$ measured in thin-film PZT (b) and PLZT (c) at room temperature and frequency of small ac field 10 kHz; (d, e) as a function of amplitude $E_{AC}$ of ac electric field measured in thin-film PZT (d, the arrow shows the first sweep of $E_{AC}$) and PLZT (e, arrow shows the sweep with maximum $E_{AC} = 1.7$ MV/m) at room temperature and frequency 10 kHz.

In epitaxial thin-film FE, the switching time at the nanosecond scale has been experimentally observed.[20] Also the possibility for the surface – stimulated nucleation of reverse domains, with an exponentially wide spectrum of waiting times, has been recently suggested.[21] Considering these findings, the fast (~ $10^{-6}$ s) switching processes could possibly contribute to the $\varepsilon'(E_{AC})$ loops. In thin-film PLZT compared to PZT, due to presence of La ions, a kind of quenched disorder can exist.[22] It can enhance a heterogeneous nucleation of FE domains and, respectively, an "accelerated" switching.[22] Somewhat broader $\varepsilon'(E_{AC})$ loops seen in thin-film PLZT were in agreement with this.

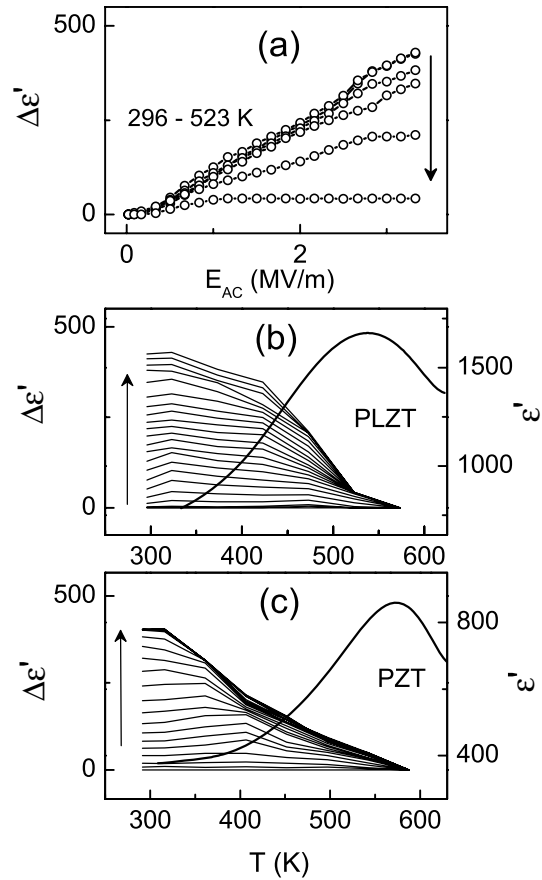

Fig. 3. The dynamic nonlinear dielectric permittivity $\Delta\varepsilon'$ (a) as a function of amplitude $E_{AC}$ of ac electric field measured in thin-film PLZT at different temperatures $T$ = 296 – 523 K (arrow shows the direction of increase of $T$); (b, c) as a function of temperature $T$ measured in thin-film PLZT (b) and PZT (c) at different amplitudes $E_{AC}$ (arrows show the direction of increase of $E_{AC}$). The temperature dependence of the permittivity $\varepsilon'$ is shown for comparison. The measurement frequency is 10 kHz.

To compare the temperature evolution of the $\varepsilon'(E_{AC})$ dependence in thin-film FE with that in thin-film RFE [Fig. 1(a)], the permittivity $\varepsilon'$ was measured using a slow increase of $E_{AC}$. Each step in $E_{AC}$ was followed by a 2 min waiting time. In thin-film PLZT, the nonlinear permittivity $\Delta\varepsilon'$ was determined as a function of $E_{AC}$ at different temperatures [Fig. 3(a)]. The nonlinear $\Delta\varepsilon'$ monotonically increased with increasing $E_{AC}$, and it decreased on heating to the temperature of the dielectric maximum, $T_m$ [Fig. 3(b)]. Above $T_m$, the nonlinearity vanished. Thin-film PZT exhibited similar type of behavior [Fig. 3(c)].

To separate contributions to the nonlinear $\Delta\varepsilon'$ either from switching or from domain wall motion, an accurate analysis of the behavior of thin-film PLZT and PZT would be necessary. Even without such an analysis it is clear that in thin-film FE, the presence of FE domains and the fast change of their configuration can be traced by $\varepsilon'(E_{AC})$ loops, monotonic increase of $\Delta\varepsilon'(E_{AC})$, and vanishing of $\Delta\varepsilon'$ above $T_m$.

In summary to this chapter, in thin films in FE state, the presence of FE domains and the fast change of their configuration determined both the dynamic dielectric nonlinearity $\Delta\varepsilon'(E_{AC})$ and the dielectric hysteresis $\varepsilon'(E_{DC})$.

### B. Thin-film relaxors

The detailed study of effect of ac electric field on the dielectric behavior in thin-film relaxor PMN has been recently performed.[6] To confirm the general character of the detected effects, thin film of another relaxor, PSN, was used in the present work.

In a PSN thin-film heterostructure, relaxor behavior was evidenced by a broad peak in the temperature dependence of the permittivity $\varepsilon'$ [Fig. 4(a)], with a frequency dispersion of the temperature $T_m$ of the dielectric maxima. The Vogel-Fulcher relationship between the temperature $T_m$ and frequency $f$ [Fig. 4(b)] was valid for the freezing temperature $T_f$ = 335 K. At high temperatures above 525 K, the inverse dielectric permittivity $1/\varepsilon'$ was found to follow the Curie-Weiss law [solid straight line in Fig. 4(c)] with the Curie constant $c \approx 2.5 \times 10^5$ K. In the temperature range between $T_m$ and 525 K, the inverse permittivity obeyed the empirical scaling law[23] $1/\varepsilon' \propto (T - T_A)^2$ [dashed curve in Fig. 4(c)], with the fitting temperature $T_A$ = 355 K. A crossover from the empirical scaling law to the Curie – Weiss law was detected at the temperature around 525 K. This made it possible to estimate the Burns temperature as approximately equal to $T_B \approx 525$ K.

At room temperature, below $T_m$, the permittivity $\varepsilon'$ exhibited a non-monotonic dependence on the amplitude $E_{AC}$ [Fig. 5(a)], similar to that observed in both epitaxial thin-film PMN[6] and polycrystalline 770 nm thick PMN film.[7] The behavior $\varepsilon'(E_{AC})$ was in agreement with the simulations.[15] Both in PSN and PMN films, the $\varepsilon'(E_{AC})$ sweeps were practically hysteresis-free, in contrast to those in thin-film FE. Applying biasing dc electric field resulted in a shift of the maximum in $\varepsilon'(E_{AC})$ to larger amplitudes [Fig. 5(b)]. The FE-like $\varepsilon'(E_{AC})$ hysteresis was not detected even at large $E_{DC}$, when an onset of FE-like state would be expected.[24]

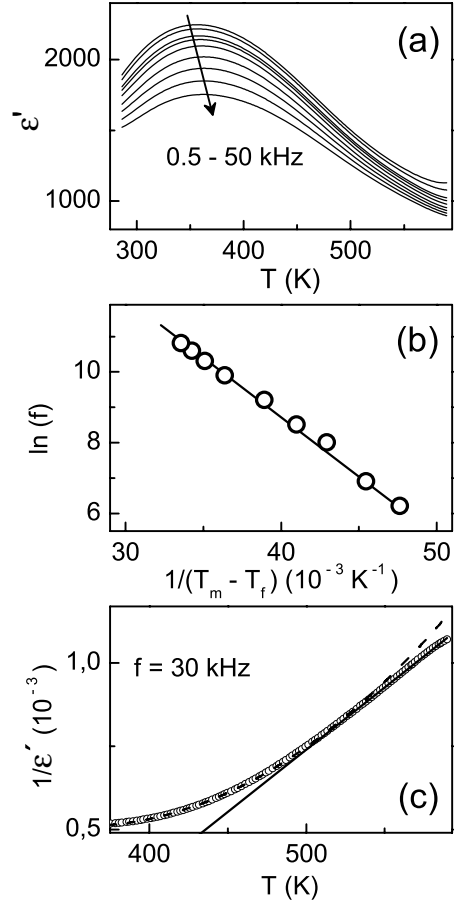

Fig. 4. The relaxor behavior in thin-film PSN. (a) The real part of the dielectric permittivity $\varepsilon'$ as a function of temperature $T$ measured on cooling at different frequencies of small ac field (the arrow shows the direction of increase of frequency). (b) The relationship between the temperature $T_m$ of the dielectric maximum and the measurement frequency $f$. The line is a fit to the Vogel-Fulcher relationship. (c) The inverse of the dielectric permittivity $1/\varepsilon'$ as a function of temperature $T$ determined at frequency 30 kHz. The straight solid line is a fit to the high-temperature Curie-Weiss behavior. The dashed curve shows a fit to the empirical scaling law.

In RFE at low temperatures, the formation of FE domains is possible.[25] In the present study in thin-film PSN, the low-temperature $P-E$ hysteresis was not obtained at fields as high as 50 MV/m, above which breakdown took place. At low temperatures, a hysteresis-free increasing $\varepsilon'(E_{AC})$ was observed [Fig. 5(c)].

Assuming presence of FE domains in thin-film PSN, such an increase of $\varepsilon'$ [curve 3 in Fig. 5(b) and Fig. 5(c)] could be explained by a subcoercive FE domain wall motion. However, the expected coercive field larger than 50 MV/m could be hardly ascribed to the thin-film FE state, in which due to presence of interfaces and defects, even a "cold" switching[21] is possible. Moreover, in thin-film FE at low temperatures [Fig. 5(d)] or under applied bias [Fig. 5(e)], the $\varepsilon'(E_{AC})$ loops did not disappear. In thin-film RFE compared to FE, different character of the $\varepsilon'(E_{AC})$ sweeps could, probably, indicate the absence of FE-like domains.

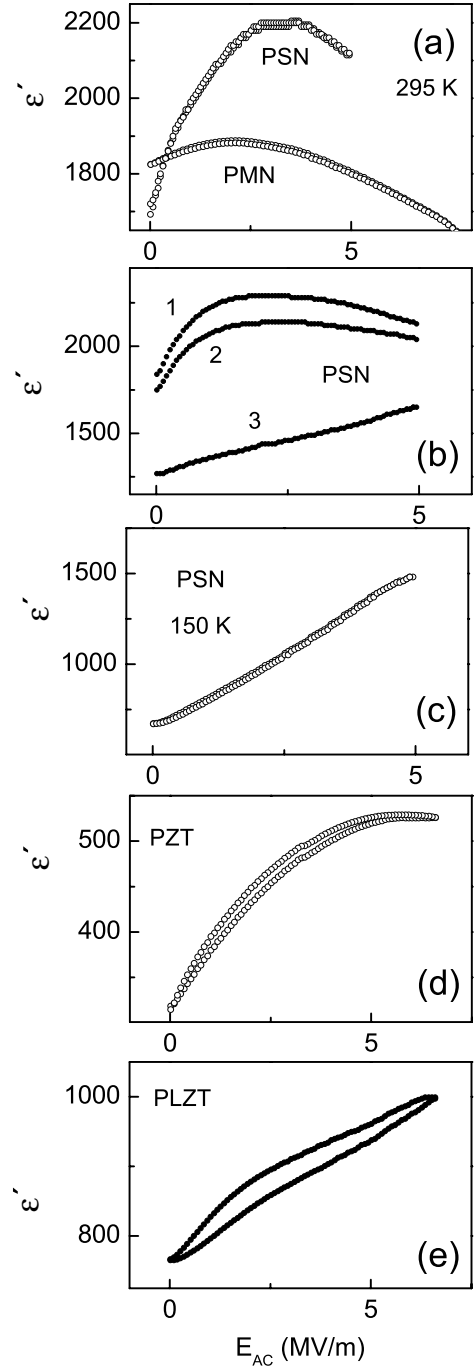

Fig. 5. The real part of the dielectric permittivity $\varepsilon'$ as a function of amplitude $E_{AC}$ of ac electric field measured at 10 kHz (a) in thin-film PMN and PSN at $T = 295$ K and $E_{DC} = 0$; (b) in thin-film PSN at $T = 295$ K and $E_{DC} = 1.25$ (curve 1), 2.5 (curve (2), and 5.0 MV/m (curve 3); (c) in thin-film PSN at $T = 150$ K and $E_{DC} = 0$; (d) in thin-film PZT at $T = 223$ K and $E_{DC} = 0$; (e) in thin-film PLZT at $T = 295$ K and $E_{DC} = 6.7$ MV/m.

In thin-film PSN, the dielectric nonlinearity $\Delta\varepsilon'$ was determined as a function of $E_{AC}$ at different temperatures [Fig. 6(a)]. Also in contrast to thin-film FE, a slower sweeping rate of $E_{AC}$ did not results in a noticeable change of the $\varepsilon'(E_{AC})$ dependence. The field dependence of the nonlinear permittivity $\Delta\varepsilon'(E_{AC})$ was in agreement with the results of numerical simulations[15] [compare Figs. 6(a) and 1(a)]. For the amplitude $E_{AC} \leq 1.75$ MV/m, the nonlinear permittivity $\Delta\varepsilon'$ increased with decreasing temperature [Fig. 6(b)], also in agreement with the simulations [Fig. 1(b)]. In contrast to thin-film FE, the permittivity $\Delta\varepsilon'$ did not vanish above $T_m$. At larger amplitudes $E_{AC} = 3 - 5$ MV/m, the permittivity $\Delta\varepsilon'$ was a nonmonotonic function of temperature [Fig. 6(c)], acquiring negative values around and above $T_m$. Also this was consistent with the simulations [Fig. 1(c)]. In thin-film PSN, the nonlinear $\Delta\varepsilon'$ was found to vanish at high temperatures above $T = 525$ K corresponding to the Burns temperature. This confirmed the model assumption that the polar entities like dipoles, different from FE domains, were responsible for polarization relaxation and dielectric nonlinearity.

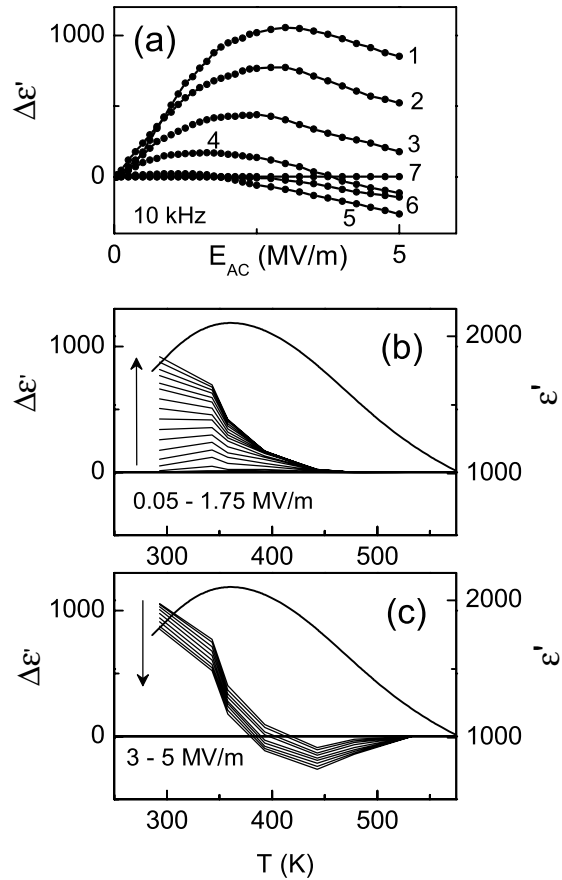

Fig. 6. The dynamic nonlinear dielectric permittivity $\Delta\varepsilon'$ measured in thin-film PSN (a) as a function of amplitude $E_{AC}$ of ac electric field at different temperatures $T = 293 - 533$ K (the curves are numbered in order of increase of $T$ with a 50 K step); (b, c) as a function of temperature $T$ at different amplitudes $E_{AC}$ (arrows show the direction of increase of $E_{AC}$). The temperature dependence of the permittivity $\varepsilon'$ is shown for comparison.

In summary to this chapter, in thin-film RFE, orientation of the randomly interacting dipoles in random field determined the dynamic dielectric nonlinearity $\Delta\varepsilon'(E_{AC})$. The presence of FE domains was not detected even at low temperatures or under applied biasing field $E_{DC}$.

### C. Relaxational hysteresis

In the dielectric behavior of thin-film RFE, no clear indications of FE domains cold be found. Respectively, FE-like switching of polarization can be hardly responsible for the $P-E$ loops observed in RFE thin films.[7-9] To trace the polarization hysteresis, the bias field dependence of the permittivity $\varepsilon'(E_{DC})$ was analyzed in more details.

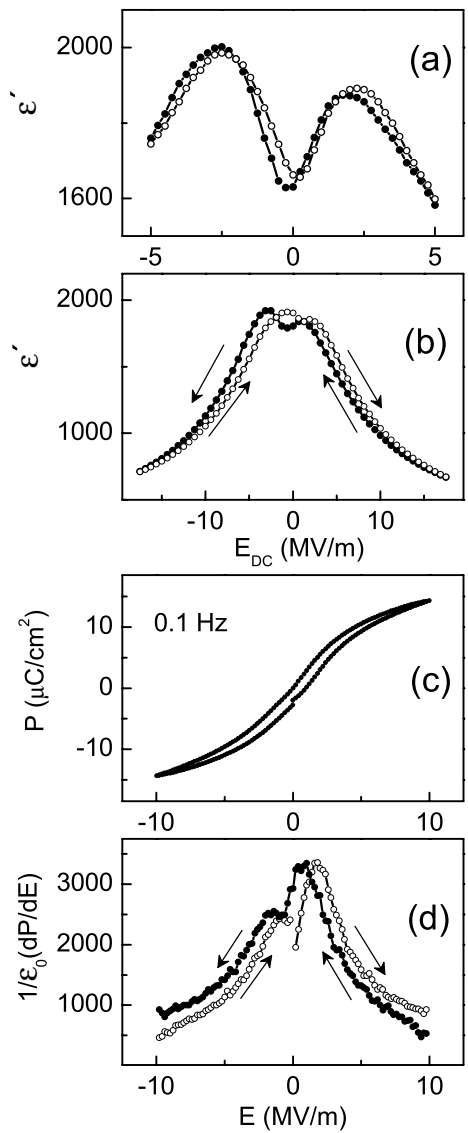

Fig. 7. Thin-film PSN. (a, b) The real part of the dielectric permittivity $\varepsilon'$ as a function of dc electric field $E_{DC}$ measured at room temperature at a sweeping rate of (a) 0.2 and (b) 0.4 MV/(m s). (c) The polarization $P$ as a function of electric field $E$ measured at room temperature at a sweeping rate corresponding to the frequency of 0.1 Hz. (d) The real part of the dielectric permittivity $\varepsilon'$ as a function of electric field obtained from the $P-E$ curve in (c).

In thin-film PSN at room temperature (below $T_m$), the permittivity $\varepsilon'$ was measured using a small amplitude $E_{AC}$ = 25 kV/m of the probing ac field and varying both the magnitude and the sweeping rate of the biasing electric field [Figs. 7(a, b)]. A considerable difference between a "weak" sweep, with the maximum $E_{DC}$ = ± 5 MV/m and the sweeping rate of about 0.2 MV/(m s) [Fig. 7(a)], and a "stronger" sweep, with the maximum $E_{DC}$ = ± 17 MV/m and the sweeping rate of about 0.4 MV/(m s) [Fig. 7(a)], was observed.

In the "weak" sweep, with increasing magnitude of $E_{DC}$, a non-monotonic change of the permittivity $\varepsilon'(E_{DC})$ resembled the ac field dependence $\varepsilon'(E_{AC})$ [compare Figs. 7(a) and 5(a)]. Such a resemblance was in a good agreement with the results of the simulations[15] considering the dc sweeps as a low-frequency ac field. The asymmetry in the $\varepsilon'(E_{DC})$ dependence was related to the asymmetry of the heterostructure and the possible presence of an in-built electric field.[26] Only a minor hysteresis of $\varepsilon'(E_{DC})$ could be detected.

In the "stronger" sweep, the dielectric hysteresis became well expressed. The double-peak structure of the $\varepsilon'(E_{DC})$ curve was smeared, but still clearly seen. Both the hysteresis and the two peaks were evidenced by the polarization measurements [Fig. 7(c)] too. The permittivity $\varepsilon'(E, f$ = 0.1 Hz) obtained by the differentiation of the $P-E$ loop [Fig. 7(d)] was consistent with the measured $\varepsilon'(E_{DC})$ and the frequency dispersion of $\varepsilon'$ below $T_m$.

In thin-film PMN, the evolution of the $\varepsilon'(E_{DC})$ dependence was studied at room temperature (above $T_m$) using both a "weak" sweep, with $E_{DC}$ = ± 1.4 MV/m and the rate 0.02 MV/(m s) [Fig. 8(a)], a "stronger" sweep, with $E_{DC}$ = ± 8 MV/m and the rate 0.2 MV/(m s) [Fig. 8(b)], and the "strongest" sweep, with $E_{DC}$ = ± 40 MV/m and the rate 1.6 MV/(m s) [Fig. 8(c)]. The behavior of $\varepsilon'(E_{DC})$ was similar to that in thin-film PSN. With increasing maximum magnitude of $E_{DC}$ and/or sweeping rate, the double-peak shape of $\varepsilon'(E_{DC})$ was smeared and disappeared. In the "strongest" sweep, the dielectric hysteresis was well resolved at the fields $E_{DC}$ = -10 – 0 MV/m. (In our previous study of thin-film PMN,[10] the dielectric hysteresis has not been noticed due to the "strong" sweeping of the field.)

The observed character of the $\varepsilon'(E_{DC})$ dependence [Figs. 7 and 8] could be qualitatively explained considering solely the contribution of the RFE dipoles, i. e., orientable, randomly interacting dipoles in random field.

The relaxation time $\tau_D$ of the orientable dipole can be estimated as[14]

$$\tau_D = \tau_0 \exp\left\{\frac{U_D - \delta(E_{INT} + E_{EXT})}{k_B T}\right\}, \qquad (2)$$

where $U_D$ is the barrier height between different dipole orientations, $E_{INT}$ is the internal electric field, $E_{EXT}$ is the external electric field, $\delta$ is a normalization coefficient, and $k_B$ is the Bolzmann constant. The distribution of the barriers $U_D$ determines a broad distribution $G(\ln\tau,T)$ of the relaxation times.

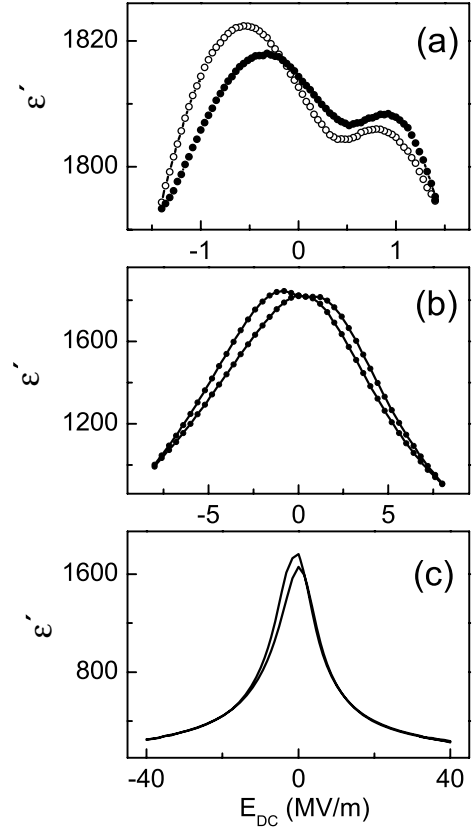

Fig. 8. Thin-film PMN. The real part of the dielectric permittivity $\varepsilon'$ as a function of dc electric field $E_{DC}$ measured at room temperature at a sweeping rate of (a) 0.02, (b) 0.2 , and (c) 1.6 MV/(m s).

The permittivity $\varepsilon'$ can be presented as

$$\varepsilon'(f,T) = \varepsilon_S(T) \int_{\tau_{min}}^{1/f} G(\ln\tau,T)d\tau , \qquad (3)$$

where $\varepsilon'_S$ is the static permittivity. Under applied field $E_{EXT}$, both the relaxation times and the permittivity change. With increasing field $E_{EXT}$, the relaxation time spectra can both move to the shorter $\tau$ and become narrower [Fig. 9(a)], leading to a non-monotonic – increasing and decreasing – field dependence of the permittivity.[6] The numerical simulations have clearly shown this effect.[15] At large enough applied field, the narrow spectrum of short relaxation times can principally be formed [Fig. 9(a)], resulting in a reduction of the frequency dispersion of the permittivity and, respectively, in "FE-looking" behavior for the studied limited frequency range [shown by the bar in Fig. 9(a)] [10]. An onset of such a polar state does not mean the appearance of FE domains.

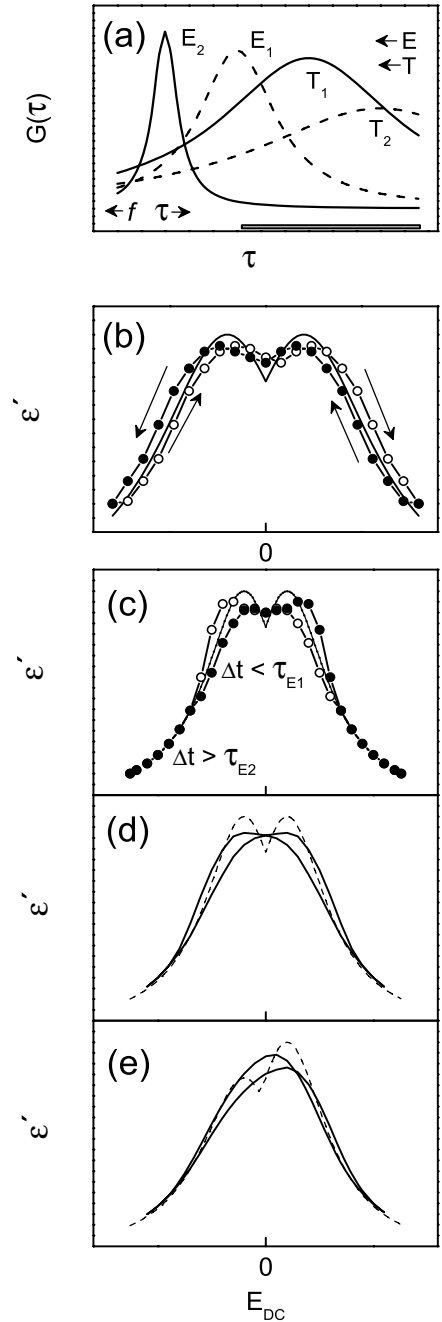

Fig. 9. (a) Schematic relaxation time spectra as a function of applied electric field $E$ and temperature $T$: for $T = T_1$ and $T = T_2$, the field $E = 0$; and for $E = E_1$ and $E = E_2$, the temperature $T = T_1$. Arrows show the direction of increase of parameter. (b – e) The real part of the dielectric permittivity $\varepsilon'$ as a function of dc electric field $E_{DC}$ determined at a relatively slow (b) and fast (c – e) sweeping rates. The hypothetic hysteresis free $\varepsilon'(E_{DC})$ is shown for comparison.

To understand the evolution of $\varepsilon'(E_{DC})$ [Figs. 7(a, b) and 8], the internal field $E_{INT}$ should be taken into account. Generally in RFE, substitutional disorder, vacancies, and displacements of ions from their equilibrium positions create the random field[14] determining $E_{INT}$. Under applied field $E_{EXT}$, the dipoles orient, and their contribution to the internal electric field changes, leading to an overall change of the internal field. This change of $E_{INT}$ requires some time, $\tau_E$. The average relaxation time $\tau_E$ of the internal field depends on both the dipole relaxation time spectrum and the applied field. The time for the complete relaxation of the internal field can be characterized by the longest relaxation time $\tau_D$ of the dipoles.

Due to a complicated connection between $E_{EXT}$, $G(\ln\tau, T)$, and $\tau_E$, the permittivity $\varepsilon'$ depends not only on the applied field $E_{DC}$, but also on the measurement time $\Delta t$ or sweeping rate. Although the study of the dependence $\varepsilon'(E_{DC}, \Delta t)$ needs a numerical analysis, some limiting cases can be qualitatively understood.

In the "weak" sweep with the relatively slow change of $E_{DC}$, the measurement time is comparable with or larger than the average field relaxation time: $\Delta t \sim \tau_E$. The permittivity $\varepsilon'$ follows the change of $E_{DC}$ without a noticeable delay time. The dielectric hysteresis cannot be detected [Figs. 7(a)].

In the "stronger" sweep, the measurement time is shorter than the field relaxation time: $\Delta t < \tau_E$. There is a delay time between applying $E_{DC}$ and a corresponding total change of the relaxation time spectrum due to a "late" change of $E_{INT}$. The permittivity $\varepsilon'$ is delayed in time from the field $E_{DC}$. For the short measurement time, this produces a "shift" of the measured $\varepsilon'(E_{DC})$ curve along $E_{DC}$-axis in the direction of the change of $E_{DC}$ [Fig. 9(b)]. The peak in $\varepsilon'(E_{DC})$ is smeared and the hysteresis can be seen [Figs. 9(c, d)]. In thin-film RFE, the observed dielectric hysteresis [Figs. 7(b) and 8(b)] was of such a relaxational nature, in contrast to the polarization switching in thin-film FE [Fig. 2(b)].

In the "strongest" sweep, the measurement time is short resulting in the relaxational hysteresis. However, at large enough $E_{DC}$ the relaxation time spectrum can become narrow and fast [Fig. 9(a)]. Then the field relaxation time is comparable with the measurement time, and the relaxational hysteresis disappears [large field in Figs. 9(c - e)]. An asymmetry of thin-film heterostructures can produce an asymmetry of the relaxational part of $\varepsilon'(E_{DC})$ [Fig. 9(e)]. In thin-film PMN, also the "strongest" sweep [Fig. 8(c)] agreed well with the suggested scenario.

With decreasing temperature, the relaxation time spectrum can move to longer $\tau$ and become broader [Fig. 9(a)]. Respectively, to obtain the time $\tau_E$ of the same order of magnitude as the measurement time, a sufficiently large field $E_{EXT}$ should be applied. This can explain, for example, why the $P-E$ loop could not be observed in thin-film PSN at low temperatures.

In summary to this chapter, in thin-film RFE, the connection between the applied field and the relaxation times of both the dipoles, $\tau_D$, and the internal field, $\tau_E$, was suggested to result in the evolution of the $\varepsilon'(E_{DC})$ dependence with respect to the magnitude and the rate of change of $E_{DC}$.

The observed dielectric hysteresis in thin-film PSN and PMN was shown to be of the relaxational nature.

### D. Dielectric hysteresis in thin-film (Ba,Sr)TiO$_3$

In thin-film BST, a coexistence of FE and RFE behavior has been demonstrated.[27] The general character of this phenomenon can be proved by many other experimental observations including both those mentioned in the Ref. 27 and those where, in our opinion, the RFE behavior has been misinterpreted as originating from the FE domain wall motion.[28] An exact reason of such coexistence remains unclear. Nevertheless, both the dynamic dielectric nonlinearity and the dielectric hysteresis, often observed above $T_m$, can be explained by the relaxorlike nature of thin-film BST.

In the previous study of Ba$_{0.4}$Sr$_{0.6}$TiO$_3$ thin-film heterostructures,[27] the Curie-Weiss behavior has been found at high temperatures above 250 K, higher than that of the dielectric maximum $T_m$ around 180 K. In the present study, the dielectric nonlinearity $\Delta\varepsilon'(E_{AC})$ of the BST film was measured as a function of amplitude $E_{AC}$ at different temperatures in the range of $T = 93 - 320$ K. The relaxorlike character of $\Delta\varepsilon'$ was clearly observed [compare Figs. 10(a) and 6(a)]. With increasing amplitude $E_{AC}$ from 0.025 to 2.0 MV/m, the permittivity $\Delta\varepsilon'$ increased with decreasing temperature [Fig. 10(b)]. For larger $E_{AC} = 2.75 - 5.0$ MV/m, it began decreasing and became negative around and above $T_m$ [Fig. 10(c)]. The nonlinear $\Delta\varepsilon'$ vanished at temperatures above 320 K, much higher than $T_m \sim 180$ K. The dynamic nonlinearity [Figs. 10(a-c)] was of the same type as that in RFE thin-film PSN [Figs. 6(a-c)], but it was completely different from that in FE thin-film PLZT and PZT [Figs. 3(b, c)].

Considering the relaxorlike dynamic dielectric nonlinearity in thin-film BST, also the room-temperature dielectric hysteresis $\varepsilon'(E_{DC})$ [Fig. 10(d)] could be explained as the relaxational one described before. It should be mentioned that in low resistivity BST films, the slow permittivity relaxation has been ascribed to a migration of oxygen vacancies.[29] However, the measurements of the transient currents performed by the same authors[30] have revealed that the vacancy related relaxation is fast (~ 0.05 s) compared to the typical measurement time in the dc sweeps. Moreover, it has been found to practically disappear on cooling from the higher temperatures to the room temperature. Although the presence of the mobile charges can affect the permittivity, it seems to be unable to determine the dielectric hysteresis.

In summary to this chapter, in thin-film BST the relaxorlike behavior was confirmed by the temperature evolution of the dynamic dielectric nonlinearity. The high-temperature dielectric hysteresis was suggested to be of the relaxational nature.

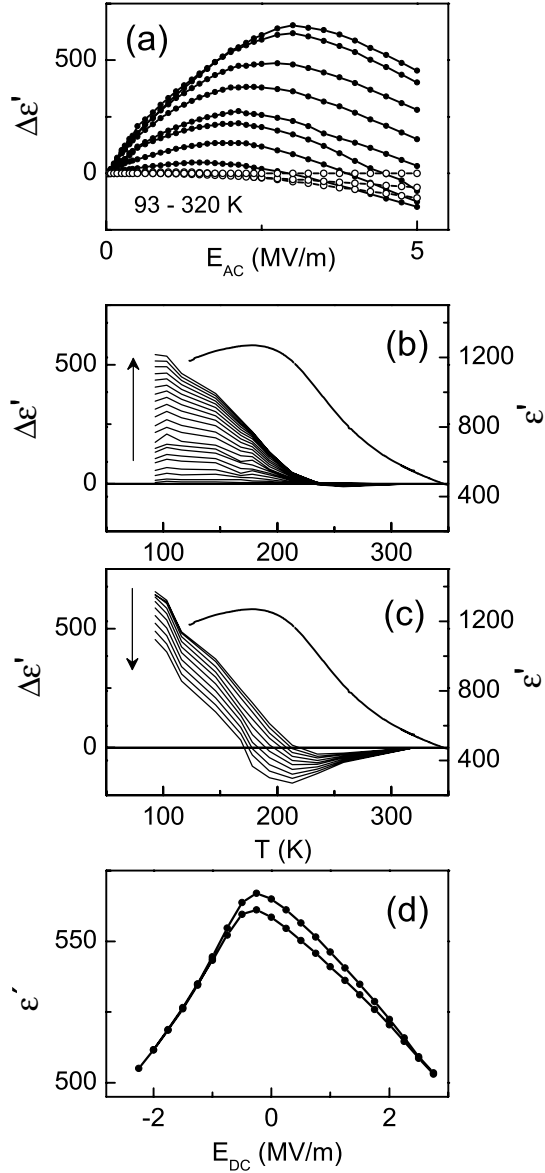

Fig. 10. Thin-film BST. (a – c) The dynamic nonlinear dielectric permittivity $\Delta\varepsilon'$ (a) as a function of amplitude $E_{AC}$ of ac electric field measured at different temperatures $T$ = 93 – 320 K (from the upper curve down); (b, c) as a function of temperature $T$ at different amplitudes $E_{AC}$ (arrows show the direction of increase of $E_{AC}$; the temperature dependence of the permittivity $\varepsilon'$ is shown for comparison). (d) The real part of the dielectric permittivity $\varepsilon'$ as a function of dc electric field $E_{DC}$ measured at room temperature.

## IV. DISCUSSION

The complicated connection between the applied electric field $E_{EXT}$, the dipole relaxation time $\tau_D$, and the field relaxation time $\tau_E$ can explain not only the results obtained here, but also many previous experimental observations. In particular, the polar state in thin-film PMN has been identified by a disappearance of the frequency dispersion of both the permittivity $\varepsilon'(f,T)$ and the temperature $T_m(f)$.[10] However, the field induced shift of the relaxation time spectra to the shorter times and the narrowing of the spectra [curves $T_1$ and $E_2$ in Fig. 9(a)] can lead to such a disappearance of the frequency dispersion even without a transition to the true FE state. Also a decrease of the freezing temperature with increasing amplitude $E_{AC}$, found in RFE ceramics and thin films,[31,6] can be explained by the same mechanism.

In the 770 nm thick films of PMN,[7] the observed nonmonotonic dependence of the effective piezoelectric coefficient on the amplitude $E_{AC}$ of the ac electric field has been ascribed to the self-polarization in the film. However, this dependence is consistent with the nonlinear behavior of the dielectric permittivity [Fig. 1(a)]. Also, both the detected relatively wide $P-E$ loops, and their shift along $E$-axis with increasing maximum applied field agree well with the relaxational mechanism of the $P-E$ hysteresis [Figs. 1(d,e)].

In the 720 nm thick film of PSN,[8] the low-temperature $P-E$ loops have been found to require an applied field by two orders of magnitude larger that that in ceramic PSN. Nevertheless, the macrodomain FE state has been suggested to exist. However, both the required very large applied field and the relatively large coercive field can be easily explained by the relaxational mechanism, without participation of fast FE domains.

In contrast to thin films in bulk RFE, neither the non-monotonic dc field dependence of the permittivity $\varepsilon'(E_{DC})$ [Figs. 7(a) and 8(a)], nor the relaxational hysteresis have been reported. In bulk RFE, the relaxation time spectrum is very broad, spreading to the range of $10^4 - 10^5$ s.[32] Respectively, the field relaxation time may be as long as $10^5$ s. Although it becomes shorter under applied field, to observe the $\varepsilon'(E_{DC})$ dependence similar to that in thin-film RFE [Fig. 7(a)], a sufficiently long measurement time should be used. During the long measurement time, the RFE state radically changes due to formation of the FE macrodomains.[24]

The main peculiarities of thin-film RFE allowing the relaxational hysteresis are the presence of the in-built electric field[26] and the broad distribution function for the random field.[33] The in-built electric field $E_{SURF}$ originates from the film-substrate mismatch. An additional polarization appears due to a piezoelectric effect arising near the surface of any film (including that with the cubic symmetry), where the symmetry inversion center is lacking. The field $E_{SURF}$ contributes to the internal field $E_{INT}$ [expression (2)], resulting in the dipole relaxation time $\tau_D$ shorter than that in the corresponding bulk RFE. On the other side, the thin-film related broadening of the random field distribution function[33] makes an onset of the long range FE order less probable compared to that in bulk RFE. In thin-film RFE, a unique combination of the "fast" dipoles and the "low-probability" domains makes it possible to observe the relaxational hysteresis.

The shape of the $\varepsilon'(E_{DC})$ hysteresis can vary depending on the dipole relaxation time spectra and the measurement conditions. The coercive field $E_C$ determined from the $P-E$ loops seems to lack a direct physical meaning. The field $E_C$ decreases with increasing measurement time, or with increasing maximum applied electric field, or with increasing temperature that makes it difficult to distinguish the RFE or FE behavior from the $P-E$ loops.

In thin-film RFE, neither the presence of the $P-E$ hysteresis loops, nor the disappearance of the frequency dispersion of the permittivity can prove the onset of the long-range FE order. The studies of the field and temperature evolution of the dynamic nonlinear dielectric permittivity $\Delta\varepsilon'(E_{AC},T)$ can partially resolve the problem. However, a direct observation of the FE domains is required. Also a theoretical analysis of both the conditions under which the FE macrodomains can be formed in epitaxial thin films of RFE and those under which the RFE state can be formed in epitaxial thin films of FE would be desirable.

## V. CONCLUSIONS

In epitaxial thin films of various perovskite RFE and FE, the dielectric permittivity was studied in the broad range of the measurement conditions.

In thin films in FE state, the presence of FE domains and the fast change of their configuration was shown to determine both the dynamic dielectric nonlinearity $\Delta\varepsilon'(E_{AC},T)$ and the dielectric hysteresis $\varepsilon'(E_{DC})$.

In thin-film RFE, orientation of the randomly interacting dipoles in random field was demonstrated to be responsible for the dynamic dielectric nonlinearity $\Delta\varepsilon'(E_{AC},T)$, while the presence of FE domains was not detected. The observed dielectric hysteresis $\varepsilon'(E_{DC})$ was suggested to be of the relaxational nature arising due to connection between the applied field and the relaxation times of both the dipoles, $\tau_D$, and the internal field, $\tau_E$.

Also in the relaxorlike thin-film $(Ba,Sr)TiO_3$, the high-temperature dielectric hysteresis was found to be of the relaxational nature.

In thin-film RFE, both the presence of the in-built electric field and the broadening of the random field distribution function were suggested to allow the relaxational hysteresis.